\documentstyle[12pt,aasms4]{article}

\lefthead{Gorgas et al.}
\righthead{Bright Spheroidal Galaxies}

\begin{document}

\title{Line-Strength Indices in Bright Spheroidals: 
Evidence for a Stellar Population Dichotomy between 
Spheroidal and Elliptical Galaxies}

\author{Javier Gorgas\altaffilmark{1}, Santos Pedraz\altaffilmark{1},
Rafael Guzm\'{a}n\altaffilmark{2}, 
Nicol\'{a}s Cardiel\altaffilmark{1}, 
J. Jes\'{u}s Gonz\'{a}lez\altaffilmark{3}}

\altaffiltext{1}{Departamento de Astrof\'{\i}sica, Facultad de
F\'{\i}sicas, Universidad Complutense, Madrid, E28040, Spain.}

\altaffiltext{2}{UCO/Lick Observatory, 
University of California, Santa Cruz, CA 95064.}

\altaffiltext{3}{Instituto de Astronom\'{\i}a, U.N.A.M., Apdo. Postal
70--264, 04510 Mexico D.F., Mexico}

\begin{abstract}
We present new measurements of central line-strength indices (namely
Mg$_2$, $\langle{\rm Fe}\rangle$, and H$\beta$) and gradients for a
sample of 6 bright spheroidal galaxies (Sph's) in the Virgo cluster.
Comparison with similar measurements for elliptical galaxies (E's),
galactic globular clusters (GGC's), and stellar population models yield
the following results: (1) In contrast with bright E's, bright Sph's are
consistent with solar abundance [Mg/Fe] ratios; (2) Bright Sph's exhibit
metallicities ranging from values typical for metal-rich GGC's to those 
for E's; (3)
Although absolute mean ages are quite model dependent, we find evidence
that the stellar populations of some (if not all) Sph's look
significantly younger than GGC's; and (4) Mg$_2$ gradients of bright
Sph's are significantly shallower than those of E galaxies. We conclude
that the dichotomy found in the structural properties of Sph and E
galaxies is also observed in their stellar populations. A tentative
interpretation in terms of differences in star formation histories is
suggested.
\end{abstract}


\section{Introduction}

Over the last few years it has been established the existence of a
structural dichotomy between elliptical (E) and spheroidal (Sph)
galaxies\footnote{Throughout this {\it Letter\/} we adopt the
nomenclature of Kormendy \& Bender\markcite{kobe} (1994), i.e.
low-density, dwarf ellipsoidal galaxies like NGC~205 are called
`spheroidals', instead of that of Binggeli\markcite{bin94} (1994), in
which these galaxies are named `dwarf ellipticals'.} 
(see e.g. Kormendy\markcite{kor85} 1985 and 
Binggeli \& Cameron\markcite{bc91} 1991) which
suggests different galaxy formation/evolution
processes for both kind of galaxies (Dekel \& Silk\markcite{desi} 1986; 
Guzm\'{a}n, Lucey \& Bower\markcite{guz} 1993).
However, this dichotomy contrasts with the remarkable similarity in the
global properties of their stellar populations. In particular, the
colour-luminosity relation as well as the correlation between the Mg$_2$
line-strength index and velocity dispersion are apparently universal for
both galaxy families (Caldwell\markcite{cal} 1983; Bender\markcite{ben}
1991). A detailed comparison of the star-formation histories of Sph's and
E's is required to provide constraints on their formation
mechanisms. Also, understanding the star-formation history of Sph's is of
vital importance for cosmological issues such as the nature of the faint
blue galaxies (Babul \& Rees\markcite{bare} 1992) or the faint end of
the luminosity function.

Recent studies devoted to compare in detail the stellar populations of
Sph's and E's have led to ambiguous results so far. For dwarf Sph's in
the Local Group, there is now clear evidence of recent (i.e., 3-5 Gyrs)
star-forming events (see e.g. da Costa\markcite{cos} 1991). 
Spectroscopic studies have provided evidence that some bright Sph's do 
exhibit a
young or intermediate age stellar population (Gregg\markcite{gre} 1991;
Held \& Mould\markcite{hemu} 1994). However, the general trend is that
bright Sph's tend to follow the galactic globular clusters (GGC's) locus
in the color--line-strength diagrams, but with a large scatter (Bothun \&
Mould\markcite{bomu} 1988; Brodie \& Huchra\markcite{brhu} 1991; Held
\& Mould\markcite{hemu} 1994). Ferguson\markcite{fer} (1994) has
compared colors and line-strengths for Sph's, E's and GGC's with the
predictions of stellar population models, concluding that differences in
the line-strength--color relations may arise due to calibration problems
and the relatively large uncertainties in both the measurements and the
population synthesis models. Clearly, a systematic study with more
precise measurements is needed to derive any serious constraints on the
stellar populations of Sph galaxies. In this {\it Letter\/} we show the
first results of a spectroscopic survey of Sph's aimed at studying in
detail their stellar populations and kinematics. The analysis presented
here is based on new central measurements of line-strength indices
(namely Mg$_2$, Fe5270, Fe5335, and H$\beta$) and, for the first
time, gradients for a sample of six Sph's in the Virgo cluster.

\section{The data}

The galaxy sample listed in Table~\ref{tbl-1} includes six bright Sph's
in the Virgo cluster. Long-slit
spectroscopic observations were carried out during April 11-15 1994, with
the $2.5m$ INT at La Palma. The
IDS spectrograph
provided 2.5 \AA (FWHM) resolution spectra in the wavelength range
4700--6100 \AA\ . The slit was aligned with the major axis. Exposure
times (typically $\sim$4 hours per galaxy) allow measurements of spectral
features out to the galaxy effective radius $r_e$. Spectra of the central
regions have very high quality (the signal-to-noise ratio ranges from 32
to 110). A detailed description of the observational setup and reduction
procedures will be given in Gorgas et al.\markcite{gor} (1997). We
emphasize that, in order to ensure reliable line-strength gradients, the
sky was carefully estimated at the slit ends taking into account any
possible contribution of the galaxy itself and the effect of scattered
light. For each galaxy spectrum we measured the Mg$_2$, H$\beta$,
Fe5270, and Fe5335 indices. The errors in these
measurements were estimated by reducing, in parallel to the galaxy
frames, error images created from photon and readout noises. Since
line-strength indices depend on spectral resolution, our spectra were
broadened to match the resolution of the widely used Lick
system (200 km s$^{-1}$). To ensure the accuracy of this correction 
and check for any
systematic errors, we observed a sample of 39 F--M stars
from the Lick stellar library (Gorgas et al.\markcite{gor93} 1993). After the 
broadening correction, we
found no systematic deviations for the Fe and H$\beta$ indices
between both data sets. The Mg$_2$ index, however, shows a systematic
offset of $0.013$ magnitudes (our values below those of Lick), which was
applied to convert our indices to the Lick system.


\section{Central line-strength indices}

In Table~\ref{tbl-1} we list, for our galaxy sample, the central Mg$_2$, 
H$\beta$ and
$\langle{\rm Fe}\rangle$ line-strength indices, and their formal errors in a
$2^{\prime\prime}\times 4^{\prime\prime}$ aperture centered on the galaxy
nucleus.
In Figure~\ref{fig1} we present line-strength diagrams for our sample of
Sph's as well as for a representative sample of GGC's and E galaxies. We 
also show the predictions of single-burst stellar population
models of a given age and metallicity (Worthey\markcite{wor} 1994).
It is
immediately apparent that, in the Mg$_2$--$\langle{\rm Fe}\rangle$ plane,
bright Sph's do not follow the extrapolation of the E sequence towards
lower Mg$_2$ values, but tend to resemble metal-rich GGC's.


The failure to reproduce theoretically the high Mg$_2$ values observed in
bright E's has been interpreted to be due to an enhancement of the
[Mg/Fe] ratio in these galaxies relative to the solar ratio assumed in
the population models (Peletier\markcite{pel} 1989; Worthey, Faber, \&
Gonz\'{a}lez\markcite{wfg} 1992; Davies, Sadler \& Peletier\markcite{dsp} 
1993, hereafter DSP). 
From Fig.~1a, it is clear that the model lines pass
through the Sph's locus thus suggesting solar abundance [Mg/Fe] ratios
for these galaxies. This difference in [Mg/Fe] between E's and Sph's may
likely imply different star-formation histories for both galaxy types. If
the enhancement of [Mg/Fe] is due to an overabundance of light elements
in bright E's, then it is
plausible that the star-formation in these galaxies occurred in a short
timescale, since Mg is created rapidly by Type II SNs whilst Fe is
produced in a longer timescale by Type Ia SNs. Under this assumption, our
observations imply a longer star-formation period in Sph's as compared to
E's. In other words, the star-forming event should have elapsed long
enough to yield a solar [Mg/Fe] ratio. However, other factors, such as
a flatter IMF in giant E's compared to Sph's, could account for the Mg
overabundance difference (Worthey et al.\markcite{wfg} 1992).

The positions of the bright Sph's in the $\langle{\rm Fe}\rangle$-- ${\rm
H}\beta$ diagram also reveal a clear dichotomy in the way Sph's and E's
populate the age-metallicity plane (Fig.~1b). Comparison between the
measured indices and the spectral synthesis models in this diagram allow us
to break the degeneracy between age and metallicity present in Fig.~1a.
Within the errors, bright Sph's
exhibit a large range of metallicities, from values typical for metal rich
GGC's ([Fe/H]$ \sim -0.75$) to those for E's.  We note that the derived
metallicities for Sph's are the same, and lower than for E's, whether
computed from Mg or Fe lines.

Concerning the age question, we find that bright Sph's span a wide range
in mean stellar ages, showing in fact a comparable age spread to that
derived, using a similar diagram, by Gonz\'{a}lez\markcite{gon} (1993,
hereafter G93) for E galaxies. Most interestingly, we find a trend 
between age
and metallicity, in the sense that ``younger'' Sph's tend to be more
metal-rich, although the sample is too small to reach a firm conclusion.
It is important to note that the computed ages are light-weighted mean
stellar ages. Moreover, since single burst models are probably a naive
approximation to the star forming history of early-type galaxies, and
given the discrepancies in the derived ages when using different models
(typically $\sim4$ Gyrs for old populations and metallicities around
solar; cf. Worthey\markcite{wor} 1994; Vazdekis et al.\markcite{alex}
1996; see Fig~1b), it is hard to give any reliable estimate of absolute
ages, but the relative trends remain.
Perhaps the key question is whether some Sph's (like UGC~7436 and
NGC~4431) are as old as the oldest E's. Fig.~1b suggests that this is the
case but this conclusion may be subject to other effects. Since the slope
of the constant-age lines (for old stellar populations) in the 
$\langle{\rm Fe}\rangle$--H$\beta$ plane is not model dependent, it is hard 
to reconcile
the relative positions of the low-H$\beta$ E's and metal-rich GGC's in
this diagram with the idea that both are old and coeval. Vazdekis et
al.\markcite{alex} (1996) have shown that the location of those
constant-age lines depends on the adopted IMF slope (see Fig~1b).
Therefore, the assumption of a flatter IMF for the bright E's would help
to explain the position of both populations (GGC's and low-H$\beta$ E's)
in this diagram without introducing important age differences, accounting
at the same time for the Mg overabundance effect.
Under this view, the stellar populations of all bright Sph's of the
sample (with solar Mg/Fe ratios) could be significantly younger than
GGC's and, therefore, than the oldest bright E's.

An important question raised by Ferguson \& Binggeli\markcite{febi} (1994) is
whether the stellar populations of Sph's resemble those in the outer regions
of E's. Using line-strength profiles from G93\markcite{gon}, when the
comparison is made at constant surface brightness ($\mu_B=20-21$ mag
arcsec$^{-2}$, corresponding to the Sph mean brightness inside our central
aperture), E galaxies attain typical Mg$_2$ line-strengths of $\sim0.28$ mag,
which are significantly larger than those observed for Sph's (see
Table~1). Therefore, the stellar populations are, again, different and local
surface brightness does not seem to be a main parameter in fixing the
properties of the stellar populations, suggesting that local stellar
populations mostly reflect the global properties (e.g. mass), and only
secondarily the local environment.

\section{Line-strength gradients}

Radial profiles of Mg$_2$, $\langle{\rm Fe}\rangle$ and H$\beta$ for our
galaxy sample are plotted in Figure~\ref{fig2}. Each
point along radius corresponds to an average of
line-strengths from symmetrical bins at both sides of the galaxy. 
For the outer regions, we co-added a
sufficient number of spectra in the spatial direction to guarantee a
minimum signal-to-noise per \AA, leading to typical
errors outside the centers of $\Delta{\rm Mg}_2=0.011$ mag,
$\Delta\langle{\rm Fe}\rangle=0.24$ \AA\ and $\Delta{\rm
H}\beta=0.36$ \AA. The gradients have been estimated from the
error-weighted linear regression fits over the radial range covering from
$1.5^{\prime\prime}$ (in order to avoid seeing effects) to the effective
radius.

As it is apparent from Fig.~2, bright Sph's possess shallow
Mg$_2$ gradients. 
The mean Mg$_2$ gradient ($\langle d{\rm Mg}_2/d\log r\rangle$)
for our sample is $-0.020$, with a r.m.s. dispersion around the
mean of $0.012$. This value can be compared with the mean Mg$_2$ gradient
derived by Gonz\'{a}lez \& Gorgas\markcite{gg} (1997) using published
and unpublished data for $109$ early-type galaxies with reliable
Mg$_2$ profiles. Following a fitting procedure similar to that described
above, the derived mean Mg$_2$ gradient for E's is $-0.055$,
with a scatter of $0.025$. Applying the Mann--Whitney
U-test, we conclude that Mg$_2$ gradients in bright Sph's are flatter
than those in E galaxies at the $0.0005$ level of confidence.
Due to their dependence on spectral resolution and photon noise, Fe
indices are worse determined than Mg$_2$. Nevertheless, Fe gradients in
bright Sph's are also found to be moderately shallow. The mean
$\langle{\rm Fe}\rangle$ gradient for the sample is $-0.32$ with
a scatter of $0.23$. Combining data from Gorgas, Efstathiou \&
Arag\'{o}n-Salamanca\markcite{gea} (1990), DSP\markcite{dsp} and 
G93\markcite{gon} we have derived a mean $\langle{\rm Fe}\rangle$ gradient
for a sample of $50$ E's of $-0.43$ (with a scatter of $0.29$).
This is also steeper than the gradients in our sample of Sph's, although
at a confidence level of only $0.26$.

Concerning H$\beta$ gradients, it is clear from Fig.~2 that
almost all bright Sph's in our sample exhibit essentially flat gradients
in the H$\beta$ strength. 
H$\beta$ line-strengths could be affected by
filling due to weak nebular emission. It must be noted, however, that
no [OIII] $\lambda 5007$ \AA\ emission lines are detectable in the
spectra of these galaxies. For our sample, we 
derive $<d{\rm
H}\beta/d\log r>=-0.11$, with a r.m.s scatter of $0.41$,
consistent with a flat mean gradient. This result agrees with what has
been found for this feature in several samples of E galaxies (Gorgas et
al.\markcite{gea} 1990; DSP\markcite{dsp}; G93\markcite{gon}).

These line-strength gradients should be interpreted in terms of radial
variations in mean age and metallicity. If we assumed that Mg$_2$
gradients are entirely due to metallicity variations within galaxies,
then, using Worthey\markcite{wor} (1994) models, these gradients would
translate into a mean $\langle\Delta[{\rm Fe/H}]/\Delta\log r\rangle =
-0.14\pm0.04$, considerably flatter than the mean
metallicity gradient, derived in a similar way, for E's ($-0.22$, Gorgas
et al.\markcite{gea} 1990; $-0.23$, DSP\markcite{dsp}; $-0.25$, Fisher,
Franx \& Illingworth\markcite{ffi} 1995). The constancy of H$\beta$
along radius in galaxies with negative metallicity gradients has been
interpreted, when taking the models literally, as an evidence for age
gradients (G93\markcite{gon}; Fisher et al.\markcite{ffi} 1995). This
result would also apply to our sample of Sph's. Although there is a large
variation from galaxy to galaxy, our data suggest moderate age gradients, 
in the sense of younger-looking stellar
populations in the inner regions relative to the outer parts. Color
gradients in Sph galaxies have previously been studied by Vader at 
al.\markcite{vvl} (1988) and Chaboyer\markcite{cha} (1994). Their main
conclusion is that Sph's exhibit shallower gradients than E's, being, in
the mean, compatible with a flat B--R colour gradient. These results are
fully consistent with our line-strength gradients since age and
metallicity effects would tend to cancel to yield shallow color
gradients. Note that, in the presence of age gradients, the above
estimate for the mean metallicity gradient would be in fact
underestimated (by $\sim0.10$ dex). In any case, since this would also
apply to giant E's, the metallicity gradients in Sph's still remain
shallower than those in E galaxies. This result suggests that SN-driven
winds, or other heating mechanism, governs the chemical evolution of
Sph's, reducing the infall of enriched gas towards the galaxy
centers and preventing, therefore, the development of steep metallicity 
gradients.

\section{Conclusions}

Through the study of absorption features, we present evidence that the
dichotomy found between the structural properties of Sph's and E's (e.g.
Bender, Burstein \& Faber\markcite{bbf} 1992) is also observed in the
stellar populations. This dichotomy rests mainly upon the observed
differences in Mg/Fe overabundance, global metallicity and metallicity
gradients. Although, in the light of the available stellar population
models, these results are still unable to identify unambiguously the
differences in the star formation histories of the E and Sph families,
our suggestion is that, while the bulk of star formation in giant E's
occurred in a short timescale and probably with a IMF skewed towards
higher masses, star formation in Sph's has proceeded, self-regulated by
galactic winds or other mechanisms, quietly in a longer timescale and
with less amount of dissipation. Further work is needed to decide whether
this elapsed star-forming period would be able to account for the young
mean ages observed in some (if not all) Sph's or whether subsequent
starbursts should be invoked. It is clear that other important aspects,
like the possible differences between the stellar populations of
nucleated and non-nucleated Sph's and the influence of environment, 
require further investigation before we can have a more
complete picture of the star-formation history in spheroidal galaxies.


\acknowledgments
We are grateful to the 
anonymous referee, A. Arag\'{o}n-Salamanca, J. Gallego and R. Peletier
for useful comments.
The INT is operated on the island of La Palma by the RGO
at the Observatorio del Roque de los Muchachos of the
Instituto de Astrof\'{\i}sica de Canarias. This work was supported in
part by the Spanish grant No. PB93-456. R. G. acknowledges funding
from the Spanish MEC fellowship EX93-27295297 and NSF grant AST91--20005.

\clearpage

\figcaption[figure1.ps]{Line-strength diagrams for GGC's (asterisks, from
Burstein et al. 1984, and Covino, Galletti \& Pasinetti 1995), E galaxies
(open circles, from G93) and the central regions of bright Sph's (filled
circles). Open triangles represent compact E's from G93 (M~32) and Gorgas
et al. (1997) (NGC~5846A and IC~767). The straight line in panel (a) is a 
least-square fit to the G93 sample of E's. Predictions from stellar population
models (Worthey 1994) are shown as dotted lines (for fixed ages of 1.5,
2, 3, 5, 8, 12 and 17 Gyr, from top to bottom in panel (b)) and
dot-dashed lines (for fixed metallicities of [Fe/H]$=-2.0, -1.5, -1.0,
-0.5, -0.25, 0.0, 0.25, 0.50$, from left to right in panel (b)). 
These lines overlap in panel (a). Full
lines in (b) represent the predictions of single-burst models from
Vazdekis et al. (1996) for a 17 Gyr old stellar population, metallicities
[Fe/H] $=-0.4, 0, 0.4$ (from left to right), and different IMF slopes
($x=1.35$ corresponds to Salpeter (1955) IMF and it is the slope used in
Worthey models). The discrepancy between both sets of models is mainly
due to differences in the temperatures of the adopted isochrones.
Error bars show the typical observational errors in the
indices of Sph's (large crosses) and E's (small crosses).
\label{fig1}}

\figcaption[figure2.ps]{Mg$_2$, $\langle{\rm Fe}\rangle$ and H$\beta$
gradients for our sample of bright Sph's. Straight lines represent
error-weighted least-squares fits to points between $1.5^{\prime\prime}$
and the effective radius ($r_e$). For each galaxy we give the derived
gradients and their formal errors. The individual profiles have been
shifted vertically by arbitrary amounts; tick marks on the ordinate axis
correspond to $0.02$ magnitudes for Mg$_2$, and 1 \AA\ for the atomic
indices ($\langle{\rm Fe}\rangle$ and H$\beta$). The central indices
listed in Table~1 correspond to the added spectra of the two innermost
points in the line-strength profiles.
\label{fig2}}

\clearpage

\begin{deluxetable}{lccccc}
\footnotesize
\tablecaption{Central line-strengths.\label{tbl-1}}
\tablewidth{0pt}
\tablehead{\colhead{Galaxy} & \colhead{M$_{\rm B}$\tablenotemark{a}} &
\colhead{\phs$\sigma$\tablenotemark{b}\phs} &
\colhead{Mg$_2$} & \colhead{H$\beta$}  
& \colhead{$\langle{\rm Fe}\rangle$\tablenotemark{c}}
}
\startdata
NGC 4415 & $-17.76$ & $-$ & 0.182$\pm$0.005 & 1.97$\pm$0.20 & 2.35$\pm$0.13 \nl   
NGC 4431 & $-17.99$ & 68  & 0.184$\pm$0.005 & 1.88$\pm$0.20 & 2.15$\pm$0.14 \nl   
NGC 4489 & $-18.76$ & 48  & 0.213$\pm$0.002 & 2.40$\pm$0.08 & 2.77$\pm$0.06 \nl   
IC  794  & $-17.42$ & 54  & 0.212$\pm$0.005 & 2.11$\pm$0.22 & 2.72$\pm$0.14 \nl   
IC 3393  & $-17.07$ & 55  & 0.142$\pm$0.006 & 2.16$\pm$0.25 & 2.13$\pm$0.17 \nl   
UGC 7436 & $-17.27$ & 45  & 0.157$\pm$0.006 & 1.90$\pm$0.27 & \phs\phd 1.88$\pm$0.18  
\tablenotetext{a}{Sources for B$_{\rm T}$ are Binggeli et al. 1985 and
Binggeli \& Cameron 1991. A distance to Virgo of $20.7$ Mpc
has been adopted (Bender et al. 1992)}
\tablenotetext{b}{Central velocity dispersion, in km s$^{-1}$, from
Bender et al. 1992 and G93}
\tablenotetext{c}{$\langle{\rm Fe}\rangle$ is defined as $({\rm
Fe5270}+{\rm Fe5335})/2$}
\enddata
\end{deluxetable}

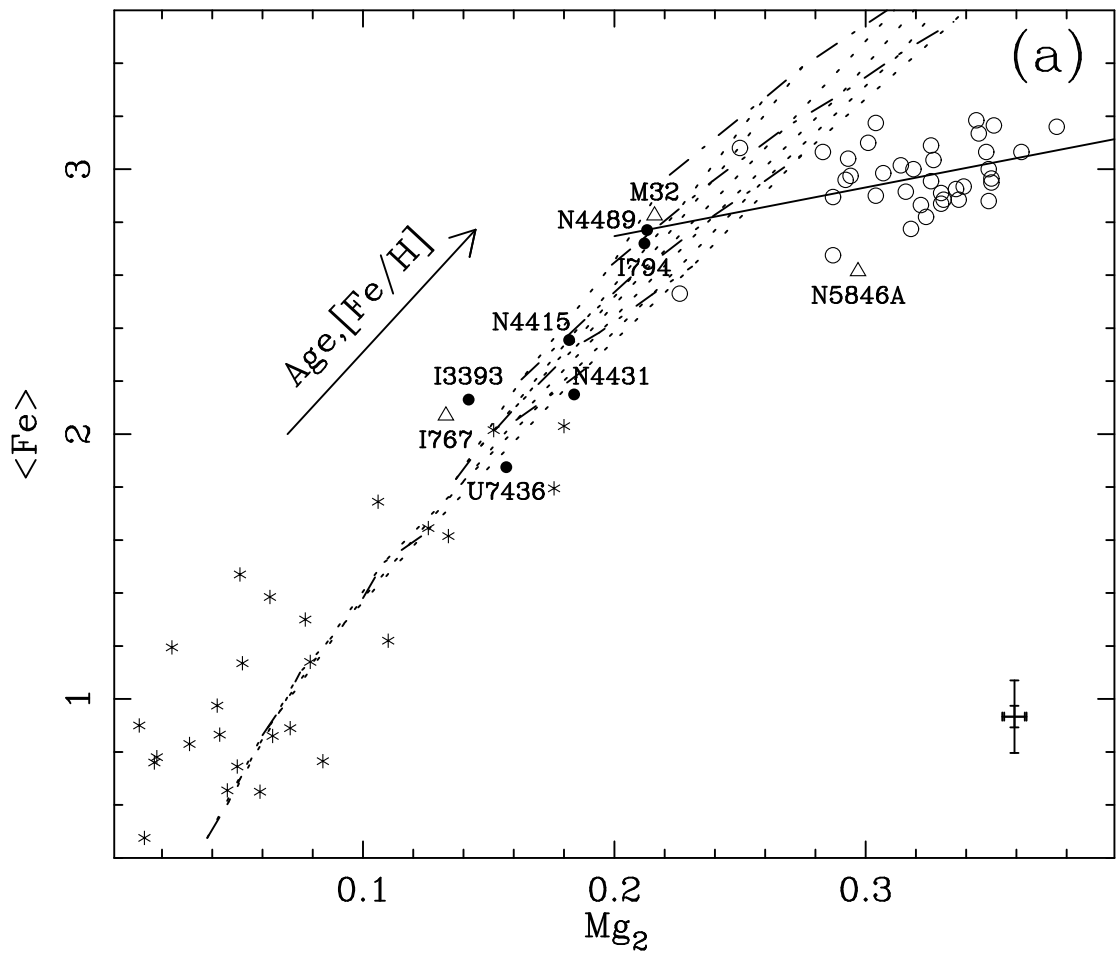
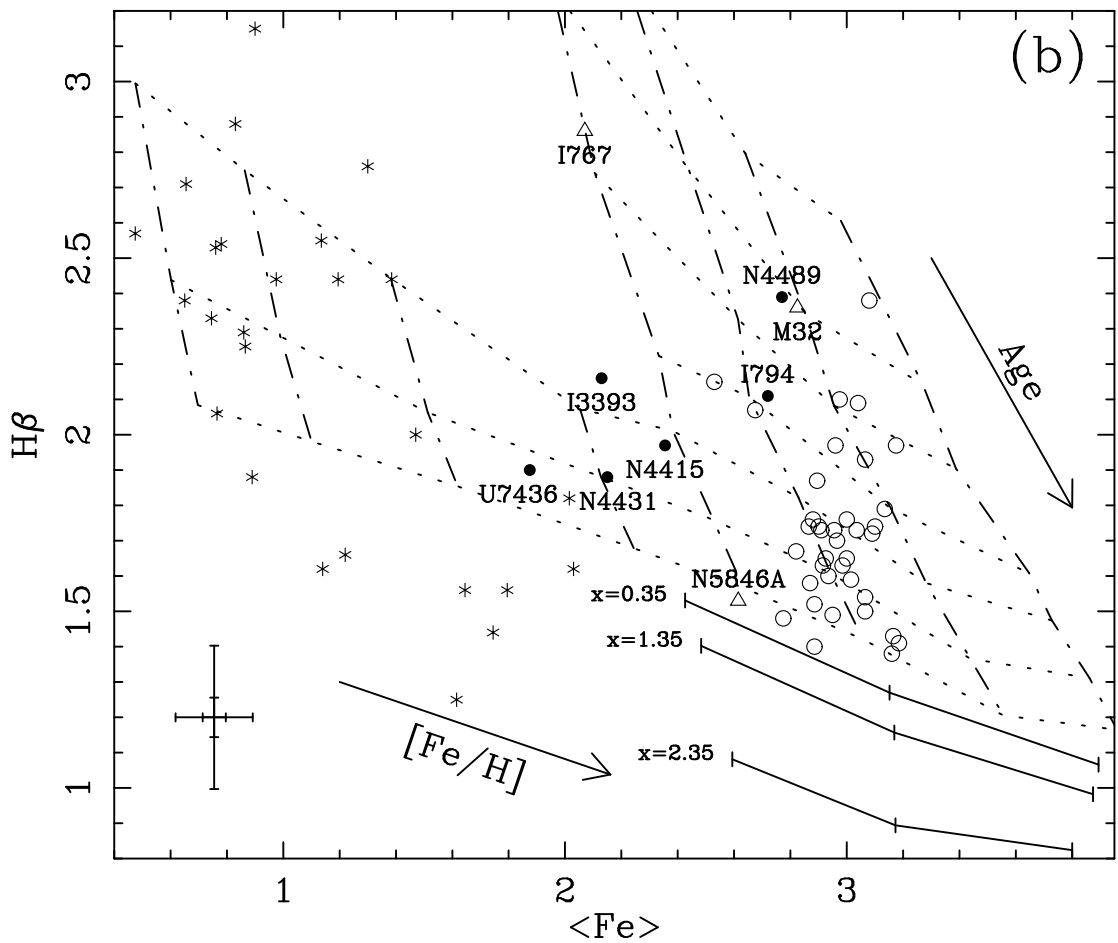
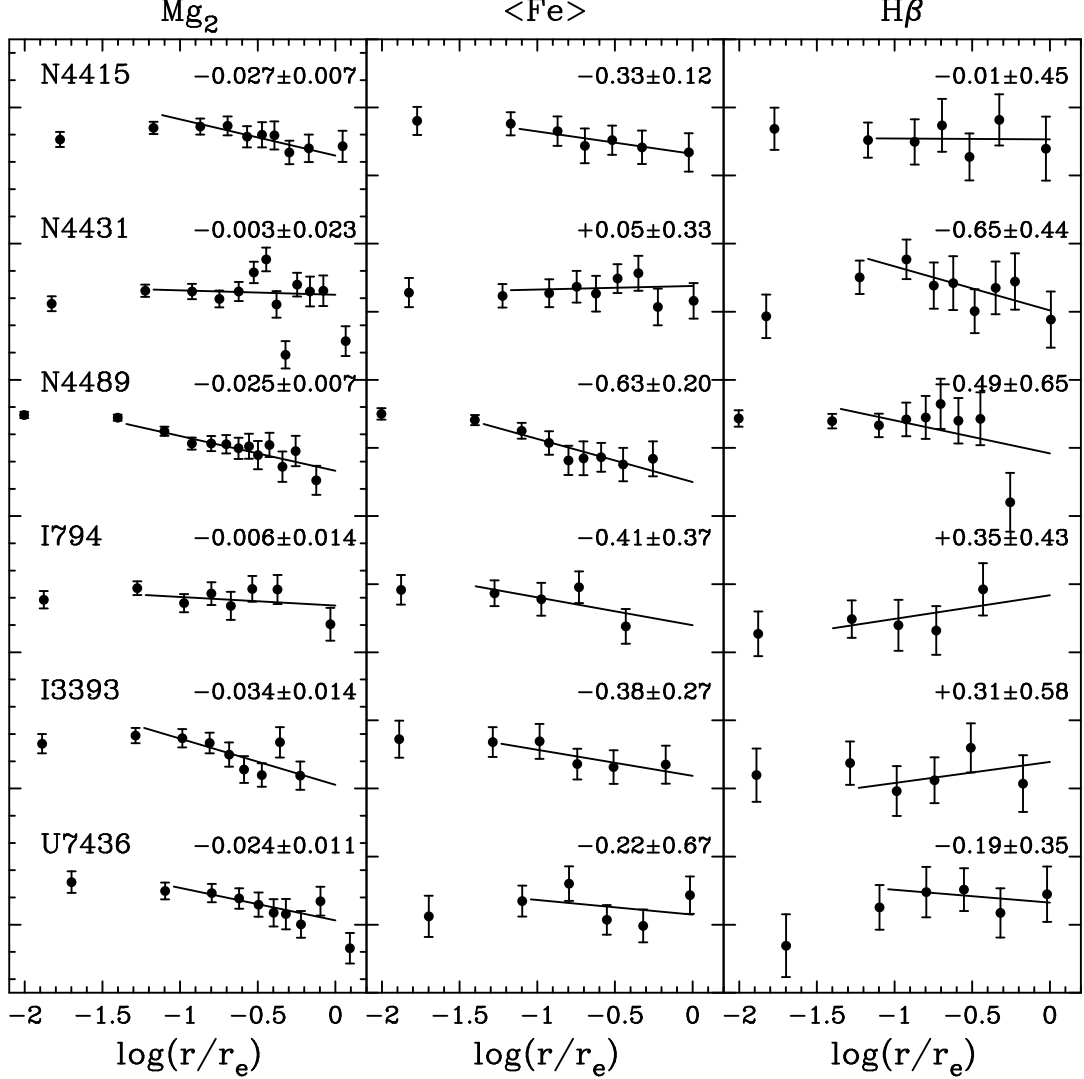
\end{document}